\documentstyle[11pt]{article}

\makeatletter

\begin{document}

\noindent{\Large {\bf About forces, }} {\Large {\bf acting on radiating
charge}}

\vspace*{5mm}\noindent {\bf Babken V. Khachatryan}\footnote{%
E-mail: saharyan@www.physdep.r.am} \newline
\noindent ${}$Department of Physics, Yerevan State University, 1 Alex
Manoogian St, 375049 Yerevan, Armenia \vspace*{10mm} \newline
\noindent {\bf Abstract.} It is shown, that the force acting on a radiating
charge is stipulated by two reasons - owing to exchange of a momentum
between radiating charge and electromagnetic field of radiation, and also
between a charge and field accompanying the charge.

\vspace*{5mm}

It is well known that the charged particle moving with
acceleration
radiates, and as a result an additional force (apart from the external one, $%
{\vec F}_0$) - force of radiation reaction acts on it. In present
paper it is shown, that this force (we shall call it as a
self-action force or simply by self-action) is a sum of two parts:
the first force is due to the exchange of the momentum between a
particle and radiation fields, i.e. the fields, which go away to
infinity. For the second force in the exchange of a momentum the
fields, accompanying a charge participate as well. These fields do
not go away to infinity, i.e. at infinity they have zero flux of
energy (details see below).

We shall start with the momentum conservation law for a system of charge and
electromagnetic field \cite{lev}, \cite{jak}

\begin{equation}
\frac d{dt}\left( {\vec P}+\frac 1{4\pi c}\int_V\left[ {\vec E\vec H}\right]
dV\right) =\frac 1{4\pi }\oint_S\left\{ {\vec E}\left( {\vec n\vec E}
\right) +{\vec H%
}\left( {\vec n\vec H}\right) -\frac{E^2+H^2}2{\vec n}\right\} dS,  \label{momcon}
\end{equation}
where ${\vec P}$ - is the particle momentum, ${\vec E}$ and ${\vec
H}$ - are the
vectors for electromagnetic field, ${\vec n}$ - is the normal to the surface $%
S$, enclosing volume $V$. On the right of formula (\ref{momcon}) the
external force ${\vec F}_0$ is omitted. From (\ref{momcon}) we can see, that
apart from external force, two forces act on the particle: force ${\vec f}_1$%
, expressed by a surface integral, and force ${\vec f}_2$ expressed by a
volume integral.

As a surface $S$ we shall take sphere of a large radius $R\rightarrow \infty
$, with the centre at the point of instantaneous place of the charge, then $%
{\vec n}={\vec R}/R$. For ${\vec E}$ and ${\vec H}$ we shall use the known
expressions for the fields created by a charged particle moving with
arbitrary velocity ${\vec v}\left( t\right) $ \cite{jak}, \cite{lanlif}

\begin{equation}
{\vec H}=\left[ {\vec nE}\right] ,\quad {\vec E}\left( {\vec r},t\right) =\frac{%
e\left( {\vec n}-{\vec \beta }\right) }{
\gamma ^2R^2x^3}+\frac e{cRx^3}\left[
{\vec n}\left[ {\vec n}-{\vec \beta },\dot{\vec \beta }\right]
\right] ,  \label{eh}
\end{equation}
where $c{\vec \beta }={\vec v}$, $\gamma =\left( 1-\beta ^2\right) ^{-\frac 12}
$, $x=1-{\vec n\vec \beta }$, $\dot{\vec \beta }\equiv d{\vec %
\beta }/dt$. Note, that all quantities in the right hand side of equation (%
\ref{eh}) are taken at the moment $t^{\prime }=t-R\left( t^{\prime }\right)
/c$.

Calculating the force $\vec f_1$ we have to substitute in
(\ref{momcon})
the term with a lowest order of $R^{-1}$ (the second term on the right in (%
\ref{eh})), corresponding, to spherical electromagnetic fields
going away to infinity, i.e. radiation fields. Then, taking into
account the remark after formula (\ref{eh}), it is possible to
write the force $\vec f_1$ in the form

\begin{equation}
{\vec f}_1=-\oint_S\frac{E^2}{4\pi }{\vec n}dS=-\oint {\vec n}\frac{dI_n}c,
\label{force1}
\end{equation}
where $dI_n$ - is the energy, radiated per unit of time in the element of
the solid angle $d\Omega $ in an arbitrary direction ${\vec n}$ \cite{lanlif}

\begin{equation}
dI_n=\frac{e^2}{4\pi cx^3}\left\{ \dot{\beta }^2+\frac 2x\left(
{\vec n}\dot{\vec \beta }\right) \left( {\vec %
\beta }\dot{\vec \beta }\right) -\frac{\left( {\vec n}%
\dot{\vec \beta }\right) ^2}{\gamma ^2x^2}\right\}
d\Omega .  \label{radiate}
\end{equation}
The formula (\ref{force1}) allows the following clear
interpretation of the origin of the force ${\vec f}_1$ : the
radiation in a direction ${\vec n}$ per unit time carries away
with itself momentum ${\vec n}dI_n/c$, and therefore, the charge
acquires a momentum $-{\vec n}dI_n/c$. As the change of a momentum
per unit time is equal to the acting force, then as a result of
radiation in a direction ${\vec n}$ the force will act on the
particle, equal to $d{\vec f}_1=$ $-{\vec n}dI_n/c$. Integrating
over all directions (over total solid angle), we get the
expression for the force ${\vec f}_1$(details for calculation see
in \cite{khac1}):
\begin{equation}
\vec f_1=-\frac Ic{\vec \beta }; \quad I=\frac{2e^2}{3c}\gamma
^4\left(
\dot{\beta }^2+\gamma ^2\left( {\vec \beta }\dot{%
\vec \beta }\right) ^2\right) .  \label{calcul}
\end{equation}
Here $I$ - is the instantaneous power of radiation, being a relativistic
invariant and having the form \cite{lanlif}, \cite{khac2}
\begin{equation}
I=-\frac 23ce^2\frac{du^k}{ds}\frac{du_k}{ds}.  \label{power}
\end{equation}
In this formula $u^k=$ $dx^k/ds$ is the four-velocity and
$ds=cdt/gamma $ is the Minkowskian interval (we follow the
notations of the book \cite{lanlif}).

Now we turn to the force ${\vec f}_2$. Here it is necessary to
take into account the contribution of both summands in formula
(\ref{eh}). The calculations are too long and, as it is easy to
see, lead to integrals, divergent at both small and long
distances. The latters are related to the divergences of the
self-energy and momentum for the point charge field. To avoid
these difficulties, we shall act as follows. Let's write a
three-dimensional equation of motion $d{\vec p}/dt={\vec f}={\vec
f}_1+$ ${\vec f}_2$ in the four-dimensional (covariant) form
\begin{equation}
\frac{dp^i}{dt}=g^i=g_1^i+g_2^i,  \label{covmotion}
\end{equation}
by entering the four-dimensional momentum $p^i=mcu^i=\left( \gamma mc,{\vec p}%
\right) $ and force $g^i=\left( \frac \gamma c{\vec f\vec \beta },\gamma /c%
{\vec f}\right) $. In formula (\ref{covmotion}) it is necessary to define $%
g_2^i$. Taking into account (\ref{calcul}) and \ref{power}, it is easy to
see, that $g_1^i$ has the form
\begin{equation}
g_1^i=\frac{2e^2}{3c}\frac{du^k}{ds}\frac{du_k}{ds}u^i.  \label{g1}
\end{equation}

As it follows from the definition of the force ${\vec f}_2$ and formula (\ref{eh}%
), where the vectors ${\vec \beta }$ and $\dot{\vec %
\beta }$ enter only, four-dimensional vector $g_2^i$ can be
expressed through the vectors $u^i$, $du^i/ds$ and $d^2u^i/ds^2$
only. The first possibility disappears as for ${\vec v}=const$,
should be $g_2^i=0$. The summand containing $du^i/ds$ is united
with a left-hand side of equation (\ref{covmotion}) and leads to
the renormalization of the charged
particle mass, so that it remains the possibility $g_2^i=\alpha d^2u^i%
/ds^2$, where $\alpha =2e^2/3c$ is a number (four-dimensional
scalar), which is determined from the requirement, that for an
arbitrary four-dimensional force $g^i$ should be $g^iu_i=0$ (to
see this it is necessary to use identity $u^iu_i=1$ and its
consequences as well). Hence
\begin{equation}
g_2^i=\frac{2e^2}{3c}\frac{d^2u^i}{ds^2}.  \label{g2}
\end{equation}
From (\ref{g2}) the expression for three-dimensional force ${\vec f}_2$
follows which we give for the reference purposes
\[
{\vec f}_2=\frac{2e^2}{3c^2}\gamma ^2\left\{ \stackrel{\cdot \cdot }{\vec %
\beta }+\gamma ^2\dot{\beta }^2{\vec \beta }+3\gamma ^2\left(
{\vec \beta }\dot{\vec \beta }\right) \dot{%
\vec \beta }+\gamma ^2\left( {\vec \beta }\ddot{\vec \beta }
\right) {\vec \beta }+4\gamma ^4\left( {\vec \beta }
\dot{\vec \beta }\right) ^2{\vec \beta }\right\} .
\]
The formulas (\ref{covmotion}), (\ref{g1}) and (\ref{g2}) lead to well-known
expression (see, for example, \cite{lanlif}) for the four-dimensional
self-action force $g^i$%
\[
g^i=\frac{2e^2}{3c^2}\gamma ^2\left( \frac{d^2u^i}{ds^2}+\frac{du^k}{ds}%
\frac{du_k}{ds}u^i\right) .
\]
Hence, for the three-dimensional self-action force ${\vec f}$ we
find (compare to the corresponding formulas in \cite{somm},
\cite{gin})
\begin{equation}
{\vec f}=\frac{2e^2}{3c^2}\left\{ {\vec A}+\left[ {\vec \beta }\left[ {\vec %
\beta \vec A }\right] \right] \right\} ,  \label{force}
\end{equation}
where ${\vec A}\equiv \gamma ^4\left( \stackrel{\cdot \cdot }{\vec \beta }%
+3\gamma ^2\left( {\vec \beta }\dot{\vec \beta }\right)
\dot{\vec \beta }\right) $.

In the nonrelativistic case $\left( \beta \ll 1\right) $, at first
approximation over $\beta $ from (\ref{force}) we get the following
expression for the self-action force (by the way we shall indicate, that
there was an error in the formula (6) in article \cite{khac2})
\begin{equation}
{\vec f}=\frac{2e^2}{3c^2}\ddot{\vec \beta }+\frac{2e^2}{c^2}%
\left( {\vec \beta }\dot{\vec \beta }\right)
\dot{\vec \beta }.  \label{force'}
\end{equation}
This force differs from the conventional one ${\vec f}^{\prime }=\frac{2e^2}{%
3c^2}\stackrel{\cdot \cdot }{\vec \beta }$, in which the essential defect is
inherent: for uniformly accelerated motion $\left( \ddot{\vec \beta }
=0\right) $, the force of radiation reaction ${\vec f}^{\prime }$
is zero, while the radiation is not equal to zero $\left( \dot{\vec
\beta }\neq 0\right) $. The force (\ref{force'}) is deprived
of this defect and always is nonzero, if the radiation is nonzero $\left(
\dot{\vec \beta }\neq 0\right) $. If $\ddot{\vec \beta }
\neq 0$ and the first summand in the right hand side of (%
\ref{force'}) dominates, then${\vec f}={\vec f}^{\prime }$; depending on the
law ${\vec \beta }\left( t\right) $, the second summand can dominate.
Generally, for $\beta \ll 1$, for self-action force it is necessary to use
the formula (\ref{force'}).

The above mentioned allows us to state that the total self-action force
acting on a radiating charge is determined by formula (\ref{force}) and it
is more appropriate to call a reaction force of radiation the force ${\vec f}%
_1$ determined by formula (\ref{calcul}). This force is always nonzero when
the particle moves with acceleration and hence radiates.

From this point of view let's consider again uniformly accelerated motion
(for arbitrary velocities). It is known that the condition for uniformly
accelerated motion has the form \cite{gin}
\begin{equation}
\frac{d^2u^i}{ds^2}+\frac{du^k}{ds}\frac{du_k}{ds}u^i=0,  \label{uniform}
\end{equation}
(thence $g^i=0$) or in three-dimensional notations
\begin{equation}
\ddot{\vec \beta }+3\gamma ^2\left( {\vec \beta }
\dot{\vec \beta }\right) \dot{\vec \beta }%
=0.  \label{uniform1}
\end{equation}
As a result for this motion the vector ${\vec A}$ goes to zero and this is
the case for the self-action force. However the radiation and radiation
reaction force are nonzero, because the acceleration is nonzero. The latter
can be easily obtained from the equation $d{\vec p}/dt={\vec F}_0+{\vec %
f}$ and is determined by the formula
\begin{equation}
mc\gamma \dot{\vec \beta }={\vec F}_0+{\vec f}-{\vec %
\beta }\left( {\vec \beta\vec F}_0\right) -{\vec \beta }\left(
{\vec \beta\vec f}%
\right) .  \label{formul}
\end{equation}
In our case for ${\vec \beta }||{\vec F}_0$, ${\vec F}_0=const$, the
acceleration is equal to
\begin{equation}
c\dot{\vec \beta }=\frac{{\vec F}_0}{m\gamma ^3}.
\label{acceler}
\end{equation}
Hence, for the uniformly accelerated motion the only force acting on charge
is the external force ${\vec F}_0$ (it can be easily checked that for the
acceleration (\ref{acceler}) the self-action force is zero). For ${\vec \beta
}\rightarrow 1$ the acceleration tends to zero, and in the case ${\vec \beta }%
\rightarrow 0$ the acceleration, as it is expected, is equal to $\frac{{\vec F%
}_0}m$.

I am grateful to the participants of the seminar of Chair of Theoretical
Physics of Yerevan State University.

\end{document}